# System Science in Politics -

# Europe and the War in Ukraine

Jürgen Mimkes, Physics Department, Paderborn University, Germany

## Abstract

Peace means order, and war brings disorder and chaos to any society. But order and disorder are not only observed in wars, in many systems they are the dominant property. Understanding order and disorder enables us to understand the structure of systems. Order and disorder are also part of the Lagrange Principle, and as statistics is valid in all systems, we may regard Lagrange statistics as a mathematical basis of system science. Two systems out of natural and social science are compared: materials of trillions of atoms and politics of millions of people.
Lagrange statistics leads to three phases of homogeneous systems: in materials we have the states: solid, liquid, gas, depending on two Lagrange parameters, temperature T (the mean energy of atoms) and pressure p. In politics we have three states: autocratic, democratic, global, depending on two Lagrange parameters, standard of living T (the mean capital of people) and political pressure p.
The three phases of each system are compared in the p-T phase diagram: Different phases of one system cannot coexist as nearest neighbors: Water will dissolve ice by exchange of atoms and heat. This leads to the present climate crisis. Democracies will dissolve autocracies by exchange of goods, ideas, and people like guest workers. This is the peaceful history of the EU and has led to the aggressive reaction of Russia in the GDR, Hungary, ČSR, and now in the Ukraine. At the end of war peaceful coexistence will not be possible between Russia and Ukraine. Only separation by a new Iron Curtain guaranteed by NATO can lead to a long-time armistice.

# 1 Introduction

To understand the war in Ukraine, one does not have to investigate Putin's head! Putin has not invented war. War is a serious disturbance of peace and general social order! War means chaos!

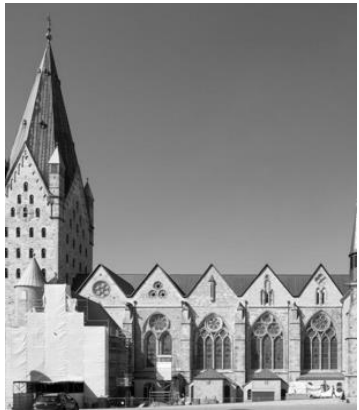 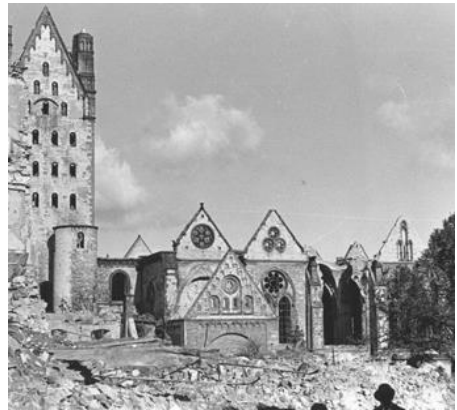

Fig. 1: Peace and order         Fig. 2: War and chaos [1]

Fig. 1 shows the Paderborn Cathedral in peace and order, Fig. 2 shows the destroyed cathedral in chaos after the Second World War. Today we see these pictures again in Ukraine, they are symbols of a destroyed order of civil society. The war in the Ukraine is a main topic in the News and TV and many authors have analyzed this war as politicians, military experts, professors of history or as local reporters [2 - 3]. These results have been based on facts, experience, meetings, or secret intelligence. This leads to the information about the present state of the Ukraine, and to the political actions of western governments [4 - 5].

This paper presents an additional channel of information, which depends only on the system and is independent of the actors; it is called systems science [6 - 17]. A system is a sum of interacting elements which are generally arranged in a special order within a certain size or volume. The elements may be real, like atoms in a crystal, or people at war, or they are abstract like the words in a book. If the elements are of the same kind, the system is called *homogeneous*, if the elements are different, the system is *heterogeneous*.

## 2. System science

The idea of system science is the existence of properties, that are the same in all systems. In this way we may transport properties from one system to another, and produce analogies founded on common mathematics. The present approach is based on the two properties common to most systems: order and disorder. A well-known example is a children's room: Even if toys initially have their neat order in the children's room , disorder will always increase during play. Only if the parents show zero tolerance for chaos, will order be maintained. Or the parents put pressure to tidy up. This example corresponds the Lagrange Principle of statistics. Joseph-Louis Lagrange (1736 - 1813) stated the principle of order (E) and disorder (S) of a system of elements in a volume (V) as: Systems will always try to take the most probable state under external constraints like order and volume,

$$L = E + T S - p V \rightarrow \text{maximum} \qquad (1).$$

The Lagrange principle consists of the functions:

1. *Lagrange function* (L): In social systems (L) represents the common happiness of a society, a society is stable if everybody is happy. In natural systems (L) corresponds to Gibbs free energy.
2. *Order* (E): Most systems have at least one kind of order that is a constraint of the system. Order can be a pattern in knitting or a behavior pattern in society. It can be a bond that creates order, e.g., by energy in crystals or by money in societies. It can be a traffic rule or a civil law.
3. *Disorder* (S): disorder is measured by entropy, the probability of errors.
4. *Volume* (V): V is another constraint, the volume taken by the elements.
5. *Lagrange-parameter* (T): The parameter (T) determines the degree of disorder. (T) is interpreted as tolerance in societies, as temperature in materials, or as standard of living in economics and politics.
6. *Lagrange parameter* (p): The value of (p) determines the size or volume available to the elements of the system and is usually interpreted as pressure (p), or as price per item in economic systems.
7. *Elements*: All functions refer to the (N) objects of the system, like numbers, atoms, people, employees, goods, bills, bites, letters, etc.

## 3. States or phases of a homogeneous system

**3.1**. The state or phase of homogeneous matter (e. g. $H_2O$) depends on two Lagrange parameters: temperature (T) and pressure (p). Matter, like $H_2O$ has three states or phases: solid (ice), liquid (water) and gas (vapor).

*Solid*: At low temperature or high pressure, $H_2O$ becomes solid ice with rigid ordered crystal bonding. Free movement of molecules is impossible.

*Liquid*: At higher temperature and low pressure, $H_2O$ becomes water with a flexible and less ordered structure. The molecules can move freely in loose bonding.

*Gas*: At very high temperature and lower pressure, $H_2O$ becomes vapor. The molecules are without bonds and move freely in space. We can now draw the corresponding phase diagram of material systems, Fig. 3.

At very high pressure, e. g. in a pressure cooker, steam becomes a liquid again even beyond 100 °C, see arrow in Fig. 3.

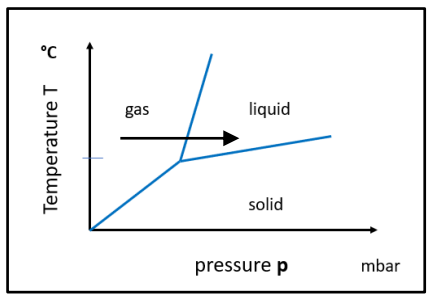 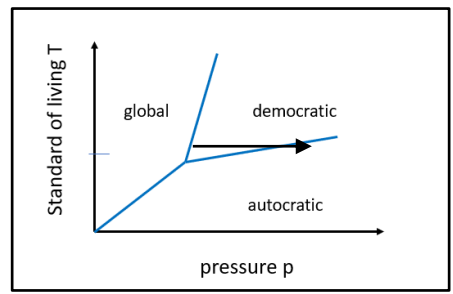

Fig. 3. Phase diagram of matter     Fig. 4. Political phase diagram

**3.2.** The political structure of a homogeneous country depends on two parameters: the standard of living (T) and internal police pressure or external military pressure (p). Politics has three different phases or states: autocratic, democratic, and global. The phase diagram of political systems in Fig. 4 is the system analogy to the phase diagram in materials, Fig. 3.

*Autocratic*: all countries in Europe are autocratic when the standard of living is very low or police or military pressure is high; a rigid and usually aggressive hierarchical order prevails. People behave collectively, there is no freedom of individual action.

*Democratic*: With a higher standard of living and less police pressure, the countries of Europe are democratic. People behave individually and can move freely with flexible local ties.

*Global*: Particularly very successful firms move almost untethered and act globally nearly without ties and taxes to any country. At high military pressure from outside like in a war, a democracy is again pressed into autocracy under martial law, see arrow in fig. 4.

Table 1 shows the GDP per capita in 2021 for many European countries.

| Country in Europe | Rank | GDP/C in US $ | Political state |
|---|---|---|---|
| Switzerland | 2 | 94,000 | Democratic |
| Sweden | 11 | 59,000 | Democratic |
| Germany | 16 | 54,000 | Democratic |
| France | 21 | 45,000 | Democratic |
| Spain | 31 | 31,000 | Democratic |
| Czechia | 37 | 25,700 | Democratic |
| Lithuania | 44 | 22,200 | Democratic |
| Poland | 49 | 18,100 | Democratic |
| Hungary | 51 | 18,000 | Democratic |
| Romania | 55 | 15,000 | Democratic |
|  |  |  |  |
| Russia | 65 | 12,000 | Autocratic |
| Belarus | 75 | 10,500 | Autocratic |
| Serbia | 87 | 8,750 | Autocratic |
| Bosnia | 95 | 7,000 | Autocratic |
| Ukraine | 116 | 4,100 | Autocratic |

Table 1. GDP (nominal) per capita 2021 [in US$] of European countries and state of government [www.imf.org/extenal/datamapper/profile/EU]

According to Table 1, countries in Europe with high mean incomes are all democratic, countries with very low mean incomes are autocratic. The transition from autocracy to democracy in Europe is currently near a value T = 15 k US$ per capita. This value is still relatively imprecise, as the level of political pressure in each country is not always known. It is only valid for closely interacting countries in Europe and may not be applicable for all countries worldwide.

## 4. Phase Diagram of Ukrainian War participants

Table 1 would lead to a more detailed political phase diagram of Europe, if the data for (p), the political or military strength (p) were given. However, these data are difficult to quantify. It is, however, possible to quantify the atomic power of USA, Russia, and Europe.

In Fig. 5 the standard of living of states involved in the Ukrainian war has been plotted as a function of military power (p) measured in nuclear war heads, as given in Fig. 6.

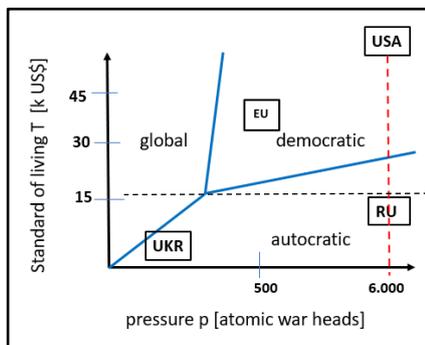

| State | GDP/C/ US $ | Nucl. heads | Polit. state |
|---|---|---|---|
| **USA** | 71.000 | 5428 | demo |
| **EU**  | 31.000 | 515  | demo |
| **RUS** | 14.000 | 5977 | auto |
| **UKR** | 5.000  | 0    | auto |

Fig. 5. Political T – p phase diagram     Fig.6. GDP/C  and military power p

The phase diagram in Fig. 5 shows an equilibrium between NATO and Russia in nuclear war heads. This has been a stabilizing factor in the danger of a nuclear war in Europe for many years and is now reducing the danger of nuclear weapons in the Ukrainian war. So far only conventional weapons have been used in the war.

## 5. EU history since 1948

**5.1.** Only at the phase borders in Fig. 3 can ice and water coexist at equal pressure and temperature. Within a phase ice and water cannot coexist. The higher temperature of water dissolves ice by exchange of heat and molecules. In nature this leads presently to the climate crisis. At the north pole the ice is melting, this leaves no space for the ice bear to live.

**5.2.** Only at the phase borders in Fig. 4 can autocracy and democracy coexist at equal pressure and standard of living. Within one phase, autocracy and democracy cannot coexist. The higher standard of living in democracies dissolves autocracies by exchange of goods, ideas, and people. After the II. World War six democracies had grown together in Europe supported by the Marshal plan in 1948. These democracies have dissolved neighboring autocracies in Spain, Portugal, Greece, and countries of the East, formerly occupied by Russia, Fig. 7. This is the history of the EU, growing peacefully from 6 to 27 states within the last 73 years. The Soviet Union has always resisted this dissolution process of the autocracy, 1953 in the DDR, 1956 in Hungary, 1968 in Czechoslovakia, etc. This process has now led to the Ukrainian war in 2022.

According to Fig. 7, the next candidates for democracy will be Belarus and Russia. Belarus has already given impressive demonstrations for democracy in 2020. However, the standard of living in Belarus is still too low to become democratic. The same it is true for Russia, and the sanctions may prevent Russia to become democratic for many more years.

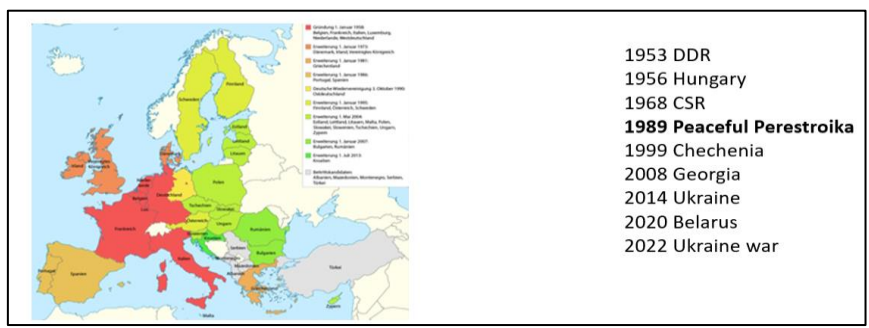

Fig. 7. Peaceful growth of the democratic EU from 6 to 27 states from 1948 to 2023, and the aggressive reaction of autocratic Russia since 1953.

## 6. Coexistence

**6.1.** There are four solutions for the coexistence for the phases ice and water in materials: At T < 0°C we have all ice and no water. At T > 0°C we have all water and no ice. Only at equilibrium temperature T = 0°C and pressure $p_0$ we have a stable coexistence of both phases. Otherwise, we must use a thermos to separate hot tea in the neighborhood of cold ice.

**6.2.** The same four solutions can be given for the coexistence of neighboring autocratic and democratic political systems:

a) Russia plans to bomb all democracies in Europe to full destruction to turn them again into autocracies like in 1945. This seems very unlikely.

b) Since 1990 Germany intensified trade with Russia, but the standard of living in Russia never reached the critical value for democracy. Trading oil or gas will not raise the standard of living of people in an autocracy, it will only support the autocrat, like in all Arabic oil countries. Only exchange of working people, everyday goods and ideas will help people to become self-assured and democratic.

c) In the phase diagram of Fig. 5 the EU and Russia are not at the same standard of living and military strength and cannot coexist, peacefully.

d) The only stable solution for coexistence left is a strict separation, a cage for the aggressor, a new Iron Curtain between Russia, and the Ukraine. This strategy has separated EU and Russia successfully from 1962 to 1989 and is still working in South Korea.

## 7. Two war strategies

There are two war strategies, where exactly the new Iron Curtain should be closed: the autocratic and the democratic strategy:

**7.1.** In the "*autocratic*" strategy each country claims to be strong enough to win the war. Russia wants to conquer all the Ukraine; their *ideal* border is at least the eastern border of Poland. The Ukraine also wants to regain the complete former territory; their *ideal* border is the old border to Russia. Both *ideals* are based on aggression, patriotism and heroes dying for their country. *Peace is obtained by victory*!

**7.2.** The "*democratic*" strategy depends on the strength of the parties Russia and Ukraine: If the Ukraine turns out to be strong enough, the western democracies will support the Ukraine to regain their former territories for a limited time. If the Ukraine equally or less strong than Russia, people in Europe will start demonstrating for peace, and the western support will have to lead to an enforced truce at a well defendable front line guarded by NATO. This new *real* border is based on defense: *Peace is obtained by negotiations*. (This democratic way also corresponds to the proposal of Henry Kissinger in the SZ of 14.1.2023, Zeit of 25.5.2023).

**7.3.** System science gives a simple answer to the question, which of these two strategies will prevail: "The diamond is an ideal crystal, it is rare and expensive. The sugar candy is a real crystal, it is common and cheap". This means politics based on ideals will rarely be successful. Only real politics with compromises will have a chance to come true.

This statement is still rather general, so far, we have only investigated homogeneous systems and their states. As a war has at least two sides, we may expect additional information, if we look at a binary system science.

## 8. Binary systems

In binary systems we have two different kinds of elements, A and B, Ukrainians, and Russians, black and white, pro and contra, Na and Cl. The Lagrange function of binary systems at constant pressure is [9]

$$L(x, T, \varepsilon) = N\{\varepsilon\, x(1-x) + T[x \ln x + (1-x)\ln(1-x)]\} \rightarrow \text{maximum !} \quad (2)$$

$$x = N_A / N; \qquad y = N_B / N = 1 - x \quad (3)$$

$$\varepsilon = E_{AB} + E_{BA} - E_{AA} - E_{BB} \quad (4)$$

$E_{AB}$ is the interaction of element A with B etc. The Lagrange function (L) depends on the relative number (x) of elements (A), the Lagrange parameter (T), and the interaction parameter epsilon ($\varepsilon$). N is the total number of elements. The interaction parameter ($\varepsilon$) characterizes three different interactions:

$\varepsilon > 0$    partnership, cooperation      (5)
$\varepsilon = 0$    integration      (6)
$\varepsilon < 0$    segregation, competition, hate, aggression      (7).

With the Lagrange function of binary systems, it is possible to model all binary interactions of nearest neighbors in a society, in materials or in other systems. A positive interaction parameter ($\varepsilon$) can model partnerships in societies or cooperation in trade. At zero the interaction parameter ($\varepsilon$) can model integration of refugees and integration into a new country. A negative interaction parameter ($\varepsilon$) models segregation of societies, competition between companies or hate and aggression in a war.

A positive interaction parameter ($\varepsilon$) models not only partnerships in societies or materials. We also model opinion formation in parliaments; obtain bicrystal with grain boundaries and point defects; binary presidential elections like in USA with party lines and dissidents; or armies at war with front lines and partisans. All these applications are equivalent systems with corresponding properties and may be used for system analysis.

## 9. Opinion formation and the end of war

On the basis of Eq. (2) we can simulate opinion formation in system science by ($\varepsilon > 0$), and we can look at four parallel processes:

**9.1.** Crystal growth of a binary crystal, NaCl, in Fig.8.

**9.2.** Simulation of crystal growth of an ordered rock salt crystal (NaCl), growing from the disordered melt according to the Lagrange principle (1) by a Monte Carlo process [17].

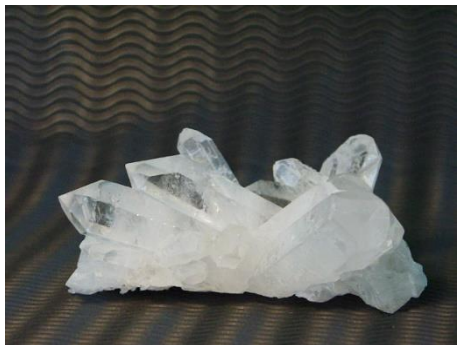  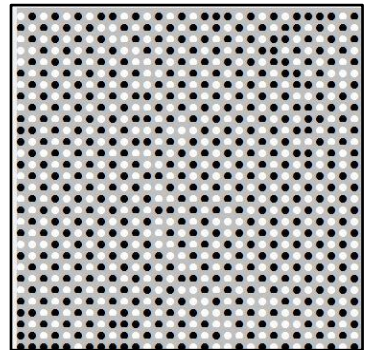

Fig. 8. Crystal with many nuclei.    Fig. 9. Na CL poly-crystal

In Fig. 9 the ordering process of NaCl takes place anywhere in the melt, Na-Cl grains will be formed as well as grains of Cl-Na. These grains will meet and form defects and finally straight grain boundaries, which may eventually lead to straight line of a bicrystal like in Fig. 8. The left side of Fig. 9 shows the crystal order with some point defects. The right side also shows point defects and in addition a curved boundary at the upper right corner. This curved line will be pushed out to the corners during the growth process, as there are more Na-Cl elements outside than Cl-Na elements inside the curve. This push-out effect may be observed in the simulation, but not in the real crystal. However, this effect may also be observed in binary elections.

**9.3** In US presidential elections each candidate determines the direction of the party. As a new candidate will slowly form the opinion of voters during the election process, it is important to have a long pre-election period.

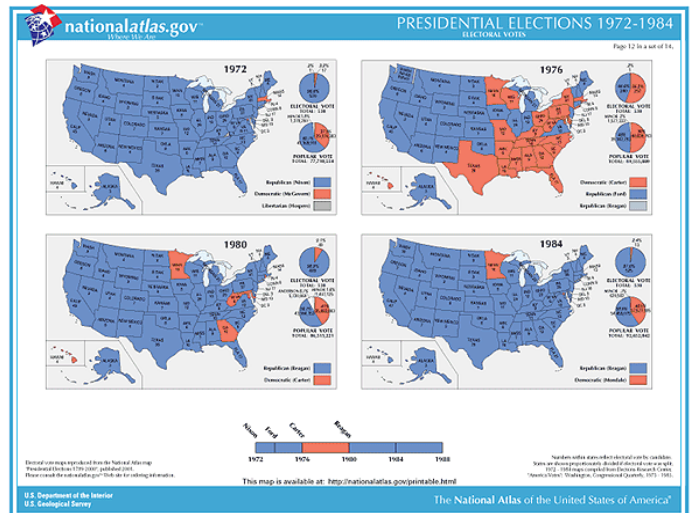

Fig. 10. US presidential elections in 1972, 1976, 1980, 1984
[www.presidency.ucsb.edu/statistics/elections]

At the end there will be boundary lines, or the winning party will push the opposition out to the corners of the map. These results have been observed in many US presidential elections, Fig. 10.

During the nearly one-year election process the losing party often was pushed. However, in all the years from 1972 to 1984 the election time was not long enough for any candidate to win all states of the country.

**9.4.** These results may be transferred to the war in Ukraine. The corresponding simulation in Fig. 9 indicates that the process of reconquering the territory occupied by Russia may take a long time. Fig. 10 indicates that it could be possible for the Ukraine to push the Russian aggressor to the corners and regain part of their former territories. However, if not, *the only chance to stop the war and the killing is a new Iron Curtain along the present fortified front lines. If Russia continues the fight, NATO weapons will have to secure this new border and freeze in the war!*

Some people claim, any compromise or freezing will reward the Russian aggression. This is true. But fairness is not part of wars, politics depends on *real* solutions. Fairness must be obtained in court later on. Lawyers have already started to cover proofs for criminal actions by Russia in the Ukraine. But courts can start working only after the war has stopped!

## 10. Outlook

Probability calculations do not contain time as a variable. Accordingly, no time schedule of the present war can be predicted by system science.

After more than one year of war peace demonstrations in European democracies are growing to end the war, see § 7.2. President Selenskyj tours all world meetings to keep western support strong. This is highly necessary. Worldwide the interest in the Ukrainian war is low outside of Europe and USA. There are only few democracies in the world, and people in autocracies have their own problems.

Fig. 11 shows a bucket of water with ice cubes floating on top. The bucket temperature is close to the melting point – near the triple point in Fig. 3.

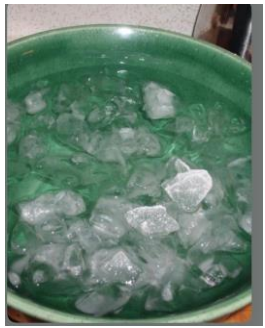  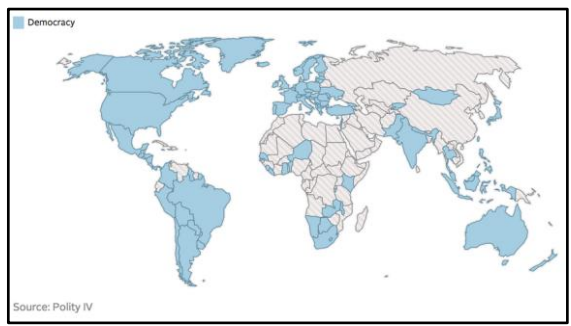

Fig. 11. Ice water          Fig. 12. Democracy (dark) in the world  [18]

In the view of System analysis, the world of democracies and autocracies in Fig. 12 corresponds to the bucket of water with floating ice cubes in Fig. 11. Revolution costs of war correspond to the latent heat of melting. Today, the mean standard of living in the connected world is about 14,000 US$/C and close to the triple point in Fig. 5. If the standard of living was divided equally, the world would be close to become democratic and peaceful. But people in poor countries can only invest in children, not in shares. This makes democracies rich and leaves large populations in Africa and Asia in poverty and autocracy. The world can only become democratic and peaceful after this problem has been solved.